# Electrical control of large magnetization reversal in a helimagnet

Yi Sheng Chai[1], Sangil Kwon[2], Sae Hwan Chun[1], Ingyu Kim[1], Byung-Gu Jeon[1], Kee Hoon Kim[1], and Soonchil Lee[2]

[1]*CeNSCMR, Department of Physics and Astronomy, Seoul National University, Seoul 151-747, Republic of Korea*

[2]*Department of Physics, KAIST, Daejeon 305-701, Republic of Korea*

**In spite of both technical and fundamental importance, reversal of a macroscopic magnetization by an electric field (*E*) has been limitedly realized and remains as one of great challenges. Here, we report the realization of modulation and reversal of large magnetization (*M*) by *E* in a multiferroic crystal $Ba_{0.5}Sr_{1.5}Zn_2(Fe_{0.92}Al_{0.08})_{12}O_{22}$, in which a transverse conical spin state exhibits a remanent *M* and electric polarization below ~150 K. Upon sweeping *E* between ± 2 MV/m, *M* is quasi-linearly varied between ± 2 $\mu_B$/f.u., resulting in the *M* reversal. Moreover, the remanent *M* shows non-volatile changes of $\Delta M$ = ± 0.15 $\mu_B$/f.u., depending on the history of the applied electric fields. The large modulation and the non-volatile two-states of *M* at zero magnetic field are observable up to ~150 K where the transverse conical spin state is stabilized. Nuclear magnetic resonance measurements provide microscopic evidences that the electric field and the magnetic field play an equivalent role, rendering the volume of magnetic domains change accompanied by the domain wall motion. The present findings point to a new pathway for realizing the large magnetization reversal by electric fields at fairly high temperatures.**

Electric field (*E*) control of magnetization (*M*) has merits over the current driven control owing to the smaller power consumption. For more than several decades, the capability has been tested in a special class of magnetic insulators, so called magnetoelectrics or multiferroics, in which *M* (or electric polarization *P*) is expected to be modulated by *E* (or magnetic field *H*)[1-3]. Such cross-coupling effects between magnetic and electric order parameters can be effectively described by direct and converse magnetoelectric (ME) effects[4]:

$$P = \alpha H + \frac{\beta H^2}{2} + ... \quad \text{and} \quad \mu_0 M = \alpha_c E + \frac{\gamma E^2}{2} + ... \quad (1)$$

where the $\alpha$ ($\alpha_c$) and $\beta$ ($\gamma$) represent the linear and quadratic ME (converse ME) coefficients. If any of the above direct ME coefficients are large, significant converse ME effects should be observable as they are roughly proportional to each other. However, most of the magnetoelectric and multiferroic compounds have so far exhibited quite small ME coefficients so that the modulation of *M* by *E* was too small to be practically useful[5-7]. It is only recent progresses that several multiferroic compounds have shown appreciable modulation of *M* with the enhanced ME coupling coefficients[8-10].

On the other hand, to exploit better magnetoelectric multiferroics in the practical devices, e.g. magnetic information storage, *M* reversal by *E* without application of a bias *H* becomes a key requirement. Toward achieving this goal, there have been very recent progresses. The reversal of ferromagnetic domain has been successfully realized in a ferromagnet/multiferroic heterostructure at room temperature (ref. 11). However, the observation is limited to interface magnetism. More recently, Tokunaga *et al.* (ref. 12) proved a bulk magnetization reversal in $Dy_{0.75}Gd_{0.25}FeO_3$ and in $Dy_{0.7}Tb_{0.3}FeO_3$ at 2.5 K, showing a maximum modulation of about ± 0.2 $\mu_B$/f.u. However, the full reversal could be only achievable at very low temperatures below 3 K, because the phenomenon stems from the coupling between rare earth and Fe moments. Meanwhile, a magnetization reversal has been demonstrated in an archetypal magnetoelectric crystal of $Cr_2O_3$ at room temperature but the reversed modulation was small ($10^{-3}$ emu/g, equivalent to ~ $3\times10^{-5}$ $\mu_B$/f.u.)[13], due to the usually small ME coefficient known in linear magnetoelectrics[5]. Therefore, it is highly desirable to search for a new mechanism that can lead to a bulk magnetization reversal with large modulation amplitude, high switching speed, and high operation temperatures.

Herein, we report the realization of modulation and reversal of large magnetization by $E$ in a multiferroic crystal $Ba_{0.5}Sr_{1.5}Zn_2(Fe_{0.92}Al_{0.08})_{12}O_{22}$, in which a transverse conical spin state exhibits a remanent $M$ and electric polarization below ~150 K. We achieved the largest reversal and modulation of $\Delta M = \pm 2$ $\mu_B$/f.u. by the electric field with the history dependent, nonvolatile changes of $\Delta M = \pm 0.15$ $\mu_B$/f.u. at zero electric field. With the nuclear magnetic resonance measurements, we found evidences that electric and magnetic domains are clamped each other, and the electric field and the magnetic field play an equivalent role of rendering the volume of magnetic domains change accompanied by the domain wall motion.

## Results

**Basic properties of the hexaferrite $Ba_{0.5}Sr_{1.5}Zn_2(Fe_{0.92}Al_{0.08})_{12}O_{22}$.** We found that the Y-type hexaferrite $Ba_{0.5}Sr_{1.5}Zn_2(Fe_{0.92}Al_{0.08})_{12}O_{22}$ (BSZFA) becomes a multiferroics with finite $P$ and $M$ and a giant ME coupling coefficient after an ME annealing procedure (Methods and Fig. 1e). BSZFA consists of a series of tetrahedral and octahedral Fe/Zn layers stacking along the $c$-axis (//[001]) (ref. 14), as shown in Fig. 1a. Under the zero field cooling (ZFC) below 60 K, it is known from the neutron diffraction studies (ref. 15) that the compound shows an incommensurate longitudinal conical ordering, in which the alternating magnetic small ($S$) and large ($L$) blocks have antiparallel $c$-axis components and in-plane helical ordering (Fig. 1b). On the other hand, in the field cooling process with in-plane 0.1 T < $\mu_0H$ < 2 T, a transverse conical state with a commensurate modulation vector $k_0 = (0, 0, 3/2)$ is stabilized (Fig. 1c, left). It is also found that the transverse conical state realized at finite $H$ still remains as a metastable phase even after removing the external $H$ (Fig. 1c, right). The material then becomes a multiferroics, exhibiting remanent $M$ and $P$ values in the $M$-$H$ (Fig. 1d) and $P$-$H$ (Fig. 1f) hysteresis loops ($P$ (//[120]) $\perp$ $M$ (//[100], Fig. 1e) and also in the $P$-$E$ hysteresis loop (Fig. 1g). In the $M$-$H$ loop, $M$ is easily saturated and changes its sign at small $H$ values of ~±30 mT with a small coercive field ~2 mT, showing that BSZFA is a very soft ferrimagnet in the transverse conical state. The $P$-$H$ hysteresis loops also show the saturation and the sign change of $P$ at about ± 40 mT, being similar to the field scale for the $M$ reversal. The overall shapes of the $M$-$H$ and $P$-$H$ loops thus look similar to each other. The observations clearly suggest that the soft ferrimagnetic domains and the associated ferroelectric domains must be clamped with each other, as similarly claimed to occur in $CoCr_2O_4$ (ref. 16).

In the $P$-$H$ hysteresis loops (Fig. 1f), it is also noted that saturated $P$ changes its sign upon changing initial poling $E_p$ direction in the ME annealing procedure (Fig. 1e). The origin of such finite $P$ with its sign changing characteristics with the $H$ bias and $E_p$ as well as the $P\perp M$ relation can be qualitatively understood via the spin current model[17,18], in which $P \sim k_0\times(\mu_S\times\mu_L)$, where $\mu_S$ and $\mu_L$ are the net moment of the adjacent $S$ and $L$ spin blocks. The sign changing behaviour of $P$ with the change of $E_p$ direction is then likely to induce different spin helicity, defined as $\Sigma\mu_S\times\mu_L$, as $E_p$ can energetically fix the local spin current direction, $\mu_S\times\mu_L$. It is also important to note that the almost linear variation of $P$ in a small $H$ bias was useful in producing the giant ME coefficient (d$P$/d$H$) of ~15000 ps/m at zero $H$ in Fig. 1f, which is the highest among the known multiferroic compounds[19]. Therefore, it is expected to observe a large converse ME effect in this compound even though the experimental test has never been made so far.

**Electric field control of *M* modulation and reversal at low temperatures.** Figure 2a demonstrates that a giant *M* modulation can be induced by the application of *E* at zero-*H* bias in the BSZFA single crystal. Before performing any experiment, the transverse conical phase was first stabilized at zero-*H* bias after performing the ME annealing with +$H_p$ (//[100]) & +$E_p$ (//[120]). Then, a slowly varying *E*//[120] was repeatedly applied between ±2 MV/m with a period of 40 sec while the magnetization *M*//[100] was recorded with a vibrating sample magnetometer measuring the magnetization in a time interval of 0.1 sec. We observed a gigantic *M* change roughly proportional to *E* with modulation of Δ*M* ~ ±2 $\mu_B$/f.u. up to *E*-sweeping of ±2 MV/m (Fig. 2a). It is noteworthy that the modulation of Δ*M* ~ ±2 $\mu_B$/f.u. is the largest magnitude ever observed in the bulk compounds. The averaged value of *M* at zero-*H* bias in this repeated measurement was approximately ~0.4 $\mu_B$/f.u. so that *M* is varied between 2.4 to −1.6 $\mu_B$/f.u., yielding unambiguously reversal of the bulk magnetization direction. The quasi-linear modulation of *M* was repeatable for many cycles of the *E*-sweeping so that the resultant *M-E* loop was quite reproducible (Fig. 2b). Interestingly, the *M-E* loops show a small hysteresis, resulting in two stable remanent *M* values ($M_r$ ~0.47 and 0.33 $\mu_B$/f.u.) at zero-*E* bias. Upon changing the ME poling into +$H_p$ & −$E_p$, the *M* shows a quasi-linear decrease with the increase of *E* (blue lines in Fig. 2a and 2b) but the full hysteresis loop maintains the same two $M_r$ values. In other words, the *M* clearly exhibits the two non-volatile $M_r$ states at zero-*E* bias, of which values depend on the history of *E*- sweep and also the sign of $E_p$ in the ME annealing procedure.

**Switch of the remanent magnetization $M_r$ by electric pulses.** Figure 2c further demonstrates that the two $M_r$ values are controllable by the applications of short *E* pulses. The left panel in Fig. 2c shows the trace of $M_r$ values, which starts at 0.6 $\mu_B$/f.u. after the ME annealing under +$H_p$ & +$E_p$ condition. Upon applying a negative short pulse of triangular *E* wave with a peak value of 6.8 MV/m and a duration time of 0.5 ms, the $M_r$ value suddenly drops to ~0.3 $\mu_B$/f.u. The first drop of $M_r$ by the $E$ pulse is relatively larger. The second negative pulse makes only negligible drop of $M_r$ while the subsequent positive pulse sharply increases $M_r$ again to ~0.45 $\mu_B$/f.u. Thus, the sequential application of the $E$ pulses of the opposite direction has resulted in sharp jumps of $M_r$ between the two independent values centred near 0.37 $\mu_B$/f.u. while the two successive $E$ pulses along the same direction do not produce appreciable $M_r$ change by the later. The switch of $M_r$ between the two values is well reproducible after many $E$ pulses without showing any fatigue effect.

Figure 2d shows the trace of $M_r$ values when the ME annealing is performed under $+H_p$ & $-E_p$ condition. In this case, the effect of the positive (negative) short pulse has been reversed but the overall behaviour is quite similar to that of the positive $E$ poling. The $M_r$ values switch between 0.35 $\mu_B$/f.u. and 0.47 $\mu_B$/f.u. with the centred value of 0.41 $\mu_B$/f.u. Such non-volatile $M_r$ changes by the short $E$ pulses imply that the ferroelectric and ferrimagnetic domain configurations at zero-$H$ bias can be tuned by the external electric field, possibly via the microscopic ME coupling inherent to the present hexaferrite. It is also noteworthy that the two nonvolatile $M_r$ values and the control between the two values by the short electric pulses can be observed even at higher temperatures.

**Electric field control of magnetization at higher temperatures.** Figure 3 summarises the $M$-$E$ hysteresis loops (Fig. 3a) and variation of the $M_r$ values by the short $E$ pulses (Fig. 3b) above 50 K. The $M$-$E$ hysteresis loop as well as the change of $M_r$ under the short $E$ pulses is clearly observable up to at least 150 K although the maximum $\Delta M$ and the magnitude of the $M_r$ jump are systematically reduced with increasing temperatures. The temperature window of 150–170 K is consistent with the region, where the phase diagram indicates the stabilization of transverse conical state only at finite magnetic fields (Supplementary Information, Fig. S1). Therefore, the clear $M$-$E$ hysteresis loop and the nonvolatile $M_r$ changes seems to occur only when the transverse conical states is stabilized at zero $H$ bias. Upon controlling the temperature for stabilizing the transverse conical states at zero $H$-bias, it is most likely that the converse ME effects and the related mechanism as found here can be also realized even at higher temperatures.

**The *M-H* loop and $^{57}$Fe NMR studies under electric field bias.** To demonstrate the change of *M* by *E* under finite *H*, we also measured the *M-H* loops under biased *E*, as shown in Fig. 1d. Overall, the magnetization curve is varying by application of the non-zero *E* bias at any given magnetic field but the relative change is close to a maximum at zero magnetic field. The coercive field under finite *E* is slightly reduced by about 10 %, and *M-H* curves slightly expand sideways. Most notably, the *M-H* curves in Fig. 1d are found to be shifted about $\mp 2$ mT by *E* of $\pm 1.2$ MV/m. The phenomena are very analogous to the usual exchange bias effect, in which the shift of the centre magnetic field occurs as if internal *H* bias is applied. In this case, the *E* does the same role of the internally applied magnetic field in the exchange bias phenomena so that the positive (negative) *E* results in the negative (positive) shift of the hysteresis loop as if the positive (negative) *H* bias is already applied. This result directly supports that the role of the *E* bias is equivalent to the *H* bias itself. Therefore, it is expected that the effect of either electric field or the magnetic field (though they are perpendicular) in this hexaferrite is equivalent, and both should change the clamped ferroelectric/ferrimagnetic domains.

To further understand the microscopic origin of such *M-H* loop changes by *E* and the large converse effects observed in Figs. 2 and 3, we systematically studied the enhancement of $^{57}$Fe NMR signal intensity. When nuclear spins are excited by an oscillating magnetic field, they usually precess along the local magnetic field. In magnetic materials, the electron spins coupled with the nuclear spins by the hyperfine interaction in a magnetic domain also precess at the same frequency. This generates oscillating magnetic field which is much stronger than that generated by the nuclear spins, resulting in the enhanced NMR signal[20]. The signal can also be enhanced in the domain wall, but by a slightly different mechanism. The oscillating magnetic field not only excites the nuclear spins but also makes the domain wall displaced back and forth at the same frequency. Then the electron spins in the domain wall oscillate, resulting in signal enhancement. Usually the enhancement effect in the domain wall is much larger than that in the domain.

The inset of Fig. 4a shows the $^{57}$Fe NMR spectra of BSZFA with and without $E$ at $\mu_0H = 0$ mT. The directions of fields are the same as that in Fig. 1e with an additional oscillating rf magnetic field $H_{rf}$ along the $ab$-plane. The NMR intensity changed under external electric and magnetic fields, while the shape and the position of the spectrum are not altered. In Figs. 4a & 4b, signal intensities vs. $H_{rf}$ are plotted for various $H$ with and without $E$ when $H_{rf} // H$ and $H_{rf} \perp H$, respectively. The intensity of the spectrum decreases overall with increasing $H$ in both cases, which is consistent with the general expectation because external magnetic field hinders the magnetic moment of electrons from vibrating freely. However, we found that the enhancement effect depends sensitively on the direction of $H_{rf}$ as summarized in Fig. 4c. The signal intensity decreases rapidly with $H$ and disappears above 50 mT when $H_{rf} // H$, whereas it is much less sensitive when $H_{rf} \perp H$. Since BSZFA is almost in the state of single magnetic domain at $\mu_0H = 50$ mT as inferred by Fig. 1d, this result suggests that the NMR signal obtained with $H_{rf} // H$ mainly comes from domain walls, and that with $H_{rf} \perp H$, domains.

The enhancement effect by the rotation of the domain magnetization becomes maximal when $H_{rf}$ and the direction of magnetization in domains are perpendicular to each other. On the other hand, the enhancement due to domain wall movement is maximal when $H_{rf}$ and the direction of magnetization in domains are parallel to each other because it makes domain walls move more easily[21]. Therefore, the fact that the NMR signal mostly comes from domains when $H_{rf} \perp H$ and from domain walls when $H_{rf}//H$ means that the direction of magnetization in domains is mostly parallel or antiparallel to $H$, and the domain walls are parallel to $H$, as depicted in Fig. 4f. The hexagonal crystal symmetry predicts in principle three easy axes in the *ab*-plane (ref. 22) but with quite small anisotropy in the plane for the transverse conical state (See, Supplementary Information Fig. S2). The magnetization of a domain is thus likely to become most parallel to $H$.

    The most striking finding in Figs. 4a & 4b is that the signal intensity changes with $E$, increasing or decreasing, depending on the presence of $H$ and polarity of $E$. Figures 4d & 4e show the $E$-dependence of the relative change of the maximal signal intensity ($\Delta I_{max}$). At $\mu_0 H = 0$ mT (Fig. 4d), $\Delta I_{max}$ decreases linearly with increasing $E$ irrespective of the direction of the electric poling ($E_p$). Since the signal intensity is proportional to the domain wall volume for $H_{rf} // H$, the decrease of $\Delta I_{max}$ indicates domain wall reduction by $E$ (Fig. 4f). The non-poled sample (black open circles) shows almost the same behaviour at $\mu_0 H = 0$ mT. At $\mu_0 H = 20$ mT (Fig. 4e), $E$ in the opposite direction to the electric poling partly recovers intensity reduced by $H$, while $E$ in the same direction reduces the intensity further. It can be also understood as change of domain wall volume: $E$ in the same direction as the poling makes the domains of which the orientation is parallel to $H$ expand, and consequently, the domain wall volume is reduced. $E$ in the opposite direction to the poling gives opposite effects. For the non-poled sample, $\Delta I_{max}$ remains almost the same irrespective of the electric field direction. This means that the half of the ferrimagnetic domain is clamped with the ferroelectric domain in the positive direction and the other half, with the one in the negative direction. This is consistent with the observation that there are two different chiralities in the conical spin ordering of BSZFA, generating two opposite directions of electric polarization. All of these results show in a microscopic level that the electric field plays an equivalent role of the magnetic field in changing the clamped ferrimagnetic/ferroelectric domain

configuration through the alignment of *P* and thus *M*, uncovering the microscopic origin of the strong cross-coupling effect realized in this hexaferrite.

It is noted in Fig. 4b that the intensity of the NMR signal from the domains behaves quite differently from that of the domain walls. The NMR signal intensity always increases in *E* irrespective of its poling direction. One possibility for this intensity change from the domain is the anisotropy field change. The shrink of coercive fields under *E* bias in the *M-H* curves in Fig. 1d implies small reduction of magnetic anisotropy field. In that case, the NMR signal is expected to be enhanced because magnetic anisotropy tends to fix the domain magnetization along a certain direction resulting in hindered magnetization rotation. Whatever the reason is, the change in the domain magnetization seems nearly irrelevant to the gigantic converse ME coupling observed in BSZFA, because the change is independent upon the polarity of applied electric field. Therefore, our NMR study shows that the gigantic, converse ME effect found in BSZFA is accompanied by the domain wall movement by electric field coming from the strong clamping of the ferroelectric and ferrimagnetic domains created in the transverse conical spin state. The NMR experiment for the zero-field cooled sample also supports this clamping picture consistently as the NMR signal intensity for the case showed no change in *E* within experimental errors. It is known that the spin structure of the zero-field cooled sample is the normal longitudinal cone, which does not induce any ferroelectricity[15].

**Fast switching speed of *M* by *E* as estimated by a GMR sensor**. The NMR results showing magnetic domain wall movement under an electric field seems unique in this insulating helimagnet, and remind of the domain wall movement by electric current in conventional metallic ferromagnets via the spin transfer torque[23]. As in the case of such spin transfer torque, the domain wall movement and resultant domain configuration changes under *E* might be also fast enough[24]. To test how fast the magnetic domain/domain wall respond to electric field, we measured the fast resistance change of a GMR sensor glued at the bottom of the crystal surface under fast triangular *E* sweep, as depicted in Fig. 5a. We find that the GMR signals vs. *E* produce almost same shape as the *M-E* loops in Fig. 2b even if the *E* sweep frequency is varied from 1 kHz to 10 kHz (Fig. 5b). When the sign of poling electric field $E_p$ is reversed, the GMR signal vs. *E* loops are also changed accordingly, very similar to the *M-E* loops in Fig. 2b. When the GMR signal vs. time is plotted together with a triangular *E* sweep applied within the time scale of ~30 μs (~10 kHz), it is clear that the curve well follows the shape of the fast sweeping of *E* (Fig. 5b). Although the speed of *E* could not be increased further due to the limit in the electronics, this result demonstrates that the switching time of *M* by *E* has an upper limit of ~30 μs and can be much smaller than 30 μs. Our observation directly shows that the speed of domain wall movement/domain configuration change by *E* can be fast enough to be useful in practical applications in this helimagnet.

**Discussion**

It is worth discussing what are the main properties the BSZFA being favourable for showing such a large converse ME effect. One of the key characteristics should be the soft ferromagnetism with very weak in-plane anisotropy realized in the transverse conical state of this hexaferrite (Fig. S2). Unlike the conical spin state with easy axis in $CoCr_2O_4$ (ref. 16), our BSZFA has the easy *ab*-plane after the field cooling process and the six-fold-hexagonal anisotropic energy then seems almost negligible inside the easy plane. The smaller the magnetic anisotropy is, the smaller the energy is likely needed to rotate *P* and *M* vectors together inside the multiferroic domains or domain walls by *E*. Moreover, as implied in Fig. S1, the transverse conical state at zero *H* in this annealed BSZFA seems stabilized up to ~150 K, below which the converse ME effects becomes significant. Therefore, the soft ferrimagnetism with minimal magnetic anisotropy and the related transverse conical spin state seems to be the most important ingredients to observe such gigantic converse ME effects. Furthermore, as other Y-type hexaferrites showing a transverse conical state at higher temperatures exist, the mechanism and phenomena as found here are likely to be applicable even at room temperature, and thus may pave a pathway for utilizing eventually the gigantic converse magnetoelectric effects in real devices at room temperature[26].

## Methods

**Single crystal growth and oxygen annealing**. The Y-type $Ba_{0.5}Sr_{1.5}Zn_2(Fe_{0.92}Al_{0.08})_{12}O_{22}$ single crystals were grown from $Na_2O$-$Fe_2O_3$ flux in air[27]. The chemicals were mixed with the molar ratio of $BaCO_3 : SrCO_3 : ZnO : Fe_2O_3 : Al_2O_3 : Na_2O$ = 2.95 : 16.74 : 19.69 : 49.32: 4.29 : 7.01 and melted at 1420 °C in a Pt crucible, followed by several thermal cycles. The single crystals of $Ba_{0.5}Sr_{1.5}Zn_2(Fe_{0.92}Al_{0.08})_{12}O_{22}$ with hexagonal shape were grown together in the same batch. The crystals were collected by checking the *c*-axis parameter using X-ray diffraction. To increase the resistivity of the specimens, an optimized heat treatment process was applied, i.e., maintaining the temperature at 900 °C under flowing oxygen condition for 8 days and cooling down at a speed of 50 °C/hour[27].

**Magnetoelectric poling procedures and magnetoelectric current experiments**. The temperature and the magnetic field were controlled using a Physical Property Measurement System (PPMS$^{TM}$, Quantum Design). In all the measurements, dc *H* was applied along the [100] direction and *E* along [120] direction. For electrical measurements, the samples were cut into a rectangular parallelepiped shape with largest surface (~1 mm$^2$) normal to the [120] direction. Before the converse ME measurements, the specimen went through the ME poling procedure starting at 120 K. For the ME poling condition of $+H_p$ & $+E_p$, the sample was electrically poled at *E* = 220 kV/m upon changing magnetic field from the paraelectric collinear ($\mu_0H$ = 2 T) to the ferrimagnetic state ($\mu_0H$ = 1.2 T) at 120 K. Then, with the same electric field and magnetic field turned on, the sample was cooled down to the lowest temperature first (15 K). *E* was then turned off at $\mu_0H = 1.2$ T, the electrodes were shorted, and the magnetic field was ramped down to zero before the converse ME experiments. For the converse ME experiments at higher temperatures, we increased temperature one by one after performing the lower temperature experiments at zero magnetic field.

The ME current was measured and integrated to determine the change of electric polarization with the magnetic field. For the ME current measurement at 15 K, almost same ME poling procedure was applied except the magnetic field was kept at 1.2 T at the final stage. The ME current was then measured by sweeping *H* either to the outside or to the inside the ferroelectric region by going across the zero field, i.e. either down to –2 T or up to +2 T or first down to –1.2 T then up to +2 T. By changing the direction of poled $E_p$, we similarly performed the same converse ME and ME current experiments under the ME poling condition of $+H_p$ & $-E_p$.

**Magnetization under electric field**. Magnetization curves were measured with a vibrating sample magnetometer in the PPMS$^{TM}$ (Quantum Design, USA). The sample was cooled down from room temperature to 15 K at $\mu_0H = 2$ T for the magnetization measurements after the field cool. For the magnetization measurements under an electric field bias, we cooled down the sample after applying the ME poling procedure starting from 120 K as described above, and the target electric field was applied at 15 K and 1.2 T before ramping down the magnetic field. Particularly for the low *H* measurement ($|\mu_0H| \leq 2$ mT), we carefully calibrated *H* of the superconducting magnet inside the PPMS with a standard paramagnetic Pd sample (Quantum Design Application Note 1070-207). For magnetization measurements under *E*, we modified the conventional VSM sample holder to allow the application of high *E*.

***P-E* loop measurement**. The *P-E* loop of the sample was measured by the PUND (positive-up-negative-down) method with our homemade setup to extract only hysteretic parts of the electric polarization change[28]. The triangular electric pulses of which duration time is 0.5 ms were applied by keeping the peak amplitude for each pulse.

**GMR sensor measurements**. The GMR chip sensor (NVE Sensors, AA004-02) is calibrated at 15 K before we apply electric field to the BSZFA sample. To work in the linear region of the GMR sensor, 2 mT bias *H* is applied. Meanwhile, 0.4 V constant excitation voltage is applied to the GMR sensor during the measurements. The configuration of the sample is chosen as such that the *M* change direction is parallel to the *H* sensitive direction of the GMR sensor.

**NMR measurements**. $^{57}$Fe NMR spectrum under bias electric fields was obtained by our homemade spectroscopy using standard spin-echo ($\pi/2$–$\pi$) sequence. The frequency of $H_1$ was 69 MHz, which is the centre frequency of the main peak[29]. The ME poling procedure for $+H_p$ & $+E_p$ is as follows: the sample was cooled down to 130 K with neither *H* nor *E*. Then the sample was cooled down to 7 K with $E$ = 250 kV/m and $\mu_0 H$ = 0.4 T.

**Acknowledgements**

This study was supported by Creative Research Initiative (2010-0018300). The work at KAIST is supported by the National Research Foundation under Grant No. NRF-2009-0078342 and 2012R1A2A2A01003598.

**Author contributions**

The first and second authors have equally contributed to this work. Corresponding authors: Kee Hoon Kim (khkim@phya.snu.ac.kr) and Soonchil Lee (soonchillee@kaist.ac.kr).

**Author contributions**

Y.S.C., S.H.C., and K.H.K. designed and initiated this work. S.H.C. grew the single crystals and Y.S.C and S.H.C carried out sample preparations for each experiment. I. K., S.H.C., Y.S.C. made the setup for VSM measurement under electric field. Y.S.C. performed VSM measurements under electric field. Y.S.C and B.-G.J. performed the GMR sensor measurements. Y.S.C. and S.H.C. performed the conventional VSM measurements. S.K. and S.L. performed the NMR measurements and discussed the data with K.H.K, S.H.C, and Y.S.C.; Y.S.C., S.H.C., S.K., S.L., and K.H.K. wrote the paper.

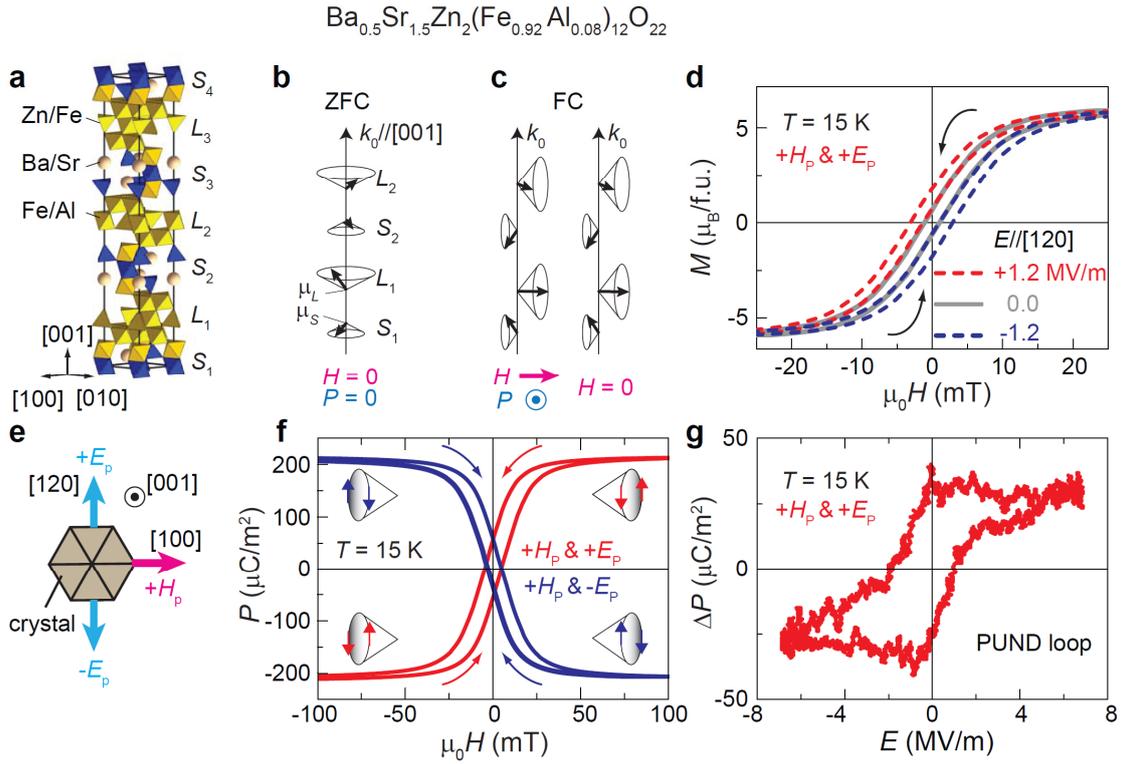

**Figure 1. Basic properties and a transverse conical state realized after application of magnetic fields in Ba$_{0.5}$Sr$_{1.5}$Zn$_2$(Fe$_{0.92}$Al$_{0.08}$)$_{12}$O$_{22}$ (BSZFA).** Schematics of (**a**), crystal structure (**b**), longitudinal conical spin state realized after zero field cool (ZFC). (**c**), transverse conical spin state realized after the field cool (FC) at 1.2 T or application of $\mu_0 H \leq 2$ T subsequent to the ZFC. $\mu_L$ and $\mu_S$ denote the net magnetic moments in the $L$ and $S$ blocks, respectively, and $k_0$ is the spin modulation vector. (**d**), *M-H*, (**f**), *P-H*, and (**g**), *ΔP-E* loops of BSZFA at 15 K. Magnetic field about 25 mT can saturate both *M-H* and *P-H* loops, leading to a single magnetic and ferroelectric domain. The spin helicity of a transverse conical state and resultant electric polarization direction are likely determined by the ME poling conditions, represented by $+H_p$ & $+E_p$ or $+H_p$ & $-E_p$ (see experimental section for procedures). The *ΔP-E* hysteresis loop based on the PUND method indicates that the sample is ferroelectric at $H = 0$ after being subject to the ME poling ($+H_p$ & $+E_p$), of which directions are schematically shown in (**e**).

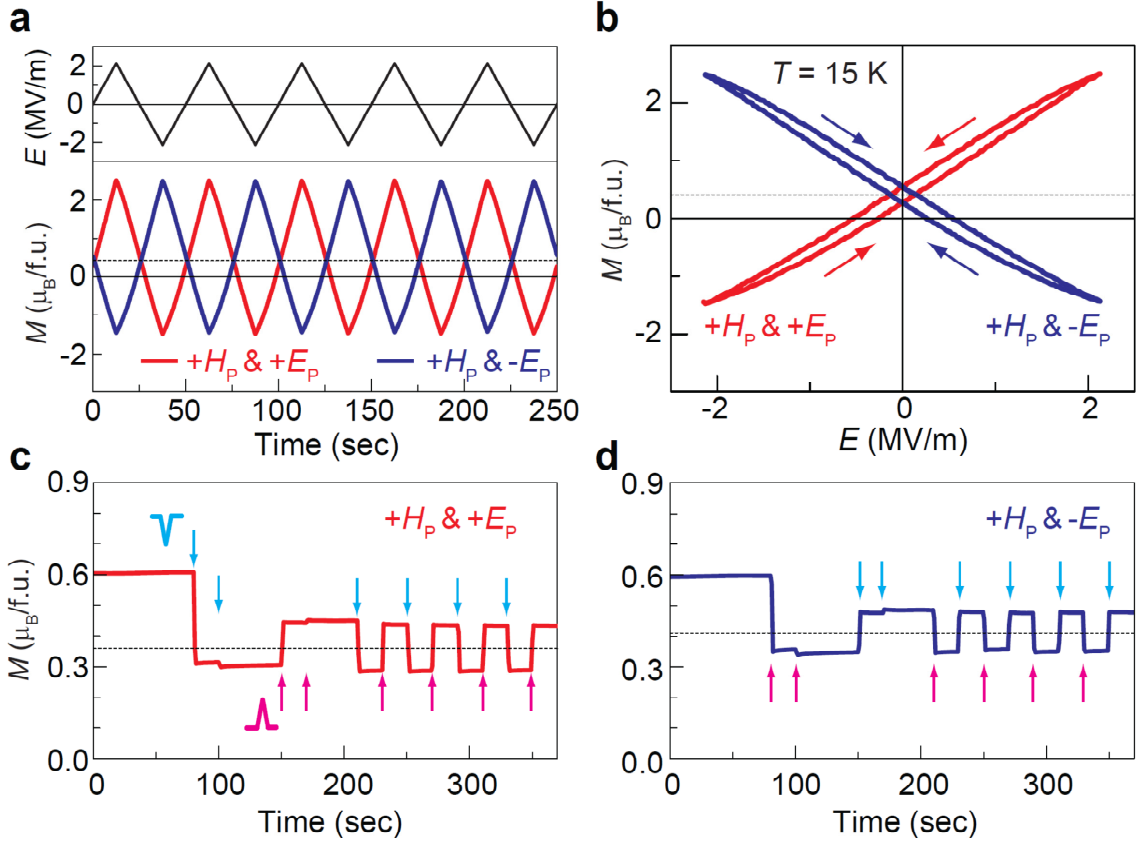

**Figure 2. Repeatable, large modulation and reversal of magnetization by electric fields.** (**a**), Periodic modulations of $M//[100]$ at a zero magnetic field under repeating triangular waves of $E//[120]$ after the ME poling. (**b**), The resultant $M$-$E$ loops showing the reversal; magnetization starts at ~0.6 $\mu_B$/f.u. due to the $+H_p$ condition and its averaged value is around 0.4 $\mu_B$/f.u. (dotted lines). (**c,d**), Jumps of the remanent $M$ by a short electric pulse with a peak field of 6.8 MV/m and duration time of 0.5 ms for the two ME poling conditions of (**c**) $+H_p$ & $+E_p$ and (**d**) $+H_p$ & $-E_p$. The magnetization, which was measured in ~0.1 second per point, seems to change immediately by the short electric pulse and the recorded information of the pulse history is non-volatile.

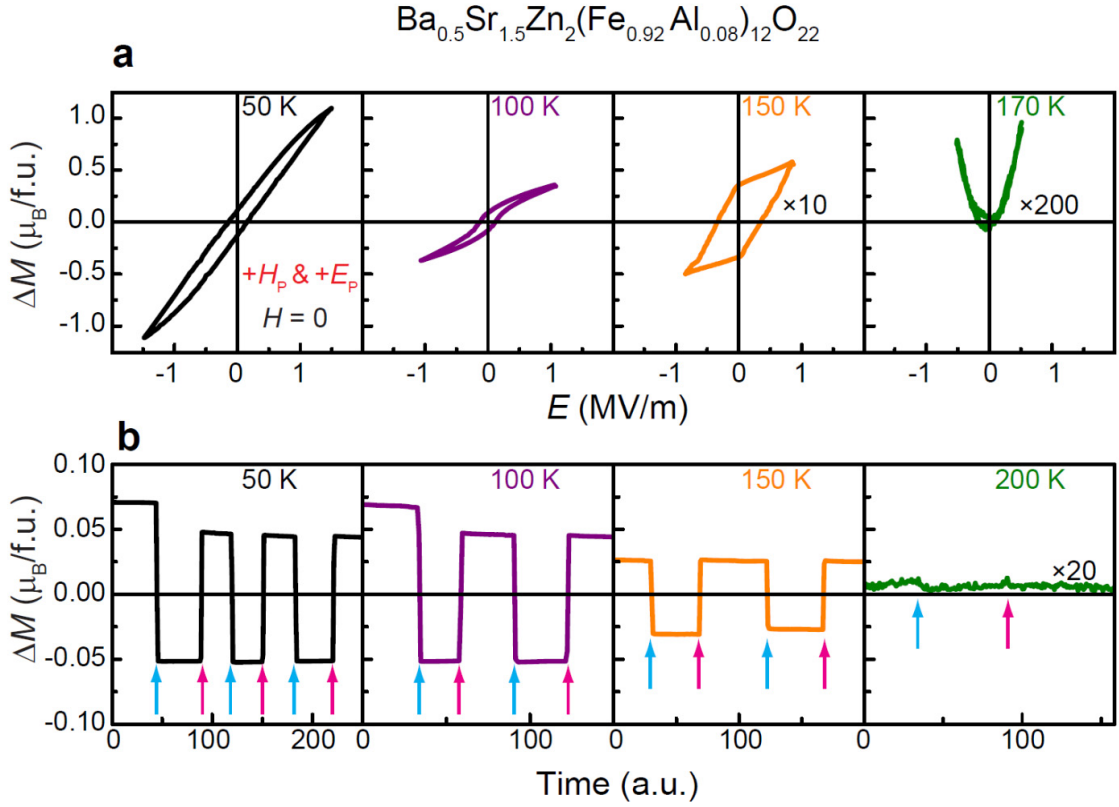

**Figure 3**. **High temperature modulation of magnetization under slow and pulsed electric fields.** (**a**), $\Delta M$-$E$ loops measured at 50 K, 100 K, 150 K, and 170 K. The hysteretic modulation was clearly observed up to 150 K (as high as 160 K). The centre magnetization value at each temperature was found to be about 0.45, 0.48, 0.10, and 0.02 $\mu_B$/f.u., respectively, which was subtracted for better displays. (**b**), The nonvolatile change of remanent $M$ after applying a triangular electric short pulse with either positive (red) or negative (blue) peak value of 4.2 MV/m, showing the $M$ jumps occur at least up to 150 K. The zero value at 200 K corresponds to the centre magnetization value of 0.03 $\mu_B$/f.u.

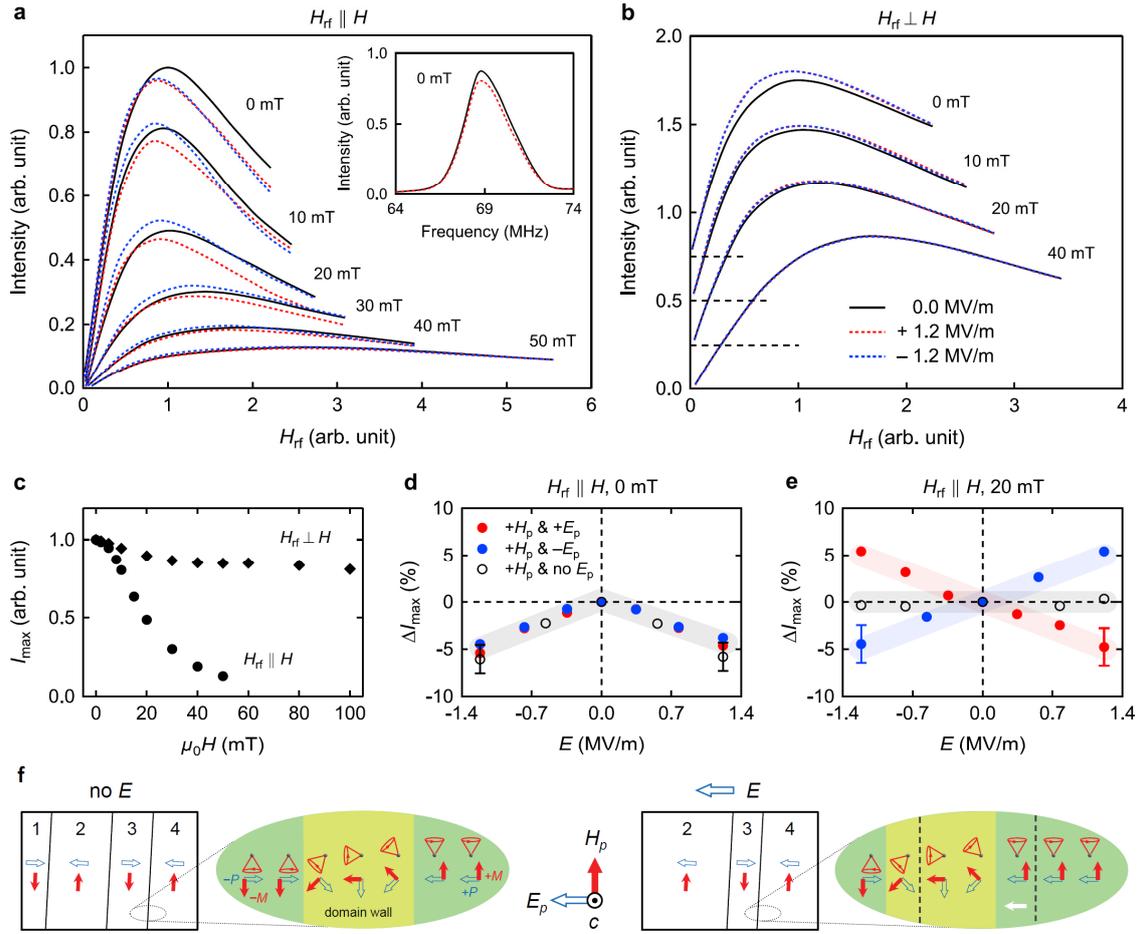

**Figure 4. The effect of magnetic and electric fields on the $^{57}$Fe NMR signal and magnetic structure of BSZFAO.** (**a,b**), The inset of **a** shows the zero field $^{57}$Fe NMR spectrum. The NMR intensity vs. $H_1$ for the + poled ($+E_p$) sample when (**a**) $H_{rf} // H$ and (**b**) $H_{rf} \perp H$. In **b**, the data from $\mu_0 H$ = 0 to 20 mT are shifted for clarity. The red dashed curves represent the data taken in the electric field of $E$ = +1.2 MV/m and the blue, $E$ = −1.2 MV/m. (**c**), $H$ dependence of the maximal signal intensity ($I_{max}$) without $E$. (**d,e**), $E$-dependence of the relative change of the maximal signal intensity ($\Delta I_{max}$) for $H_{rf} // H$ at 0 and 20 mT. The red and blue solid circles were taken after $+E_p$ and $-E_p$ electric poling, respectively. The black open circles were taken without ME poling. All data were obtained at 7 K. (**f**), Schematic diagrams of domain configuration under external $E$ and $H$. The two types of the clamped ferroelectric and ferrimagnetic (multiferroic) domains hold inside an orthogonal $M$ and $P$ relationship, and their domain walls move under external $E$ and thus lead to the volume increase of one type of multiferroic domains.

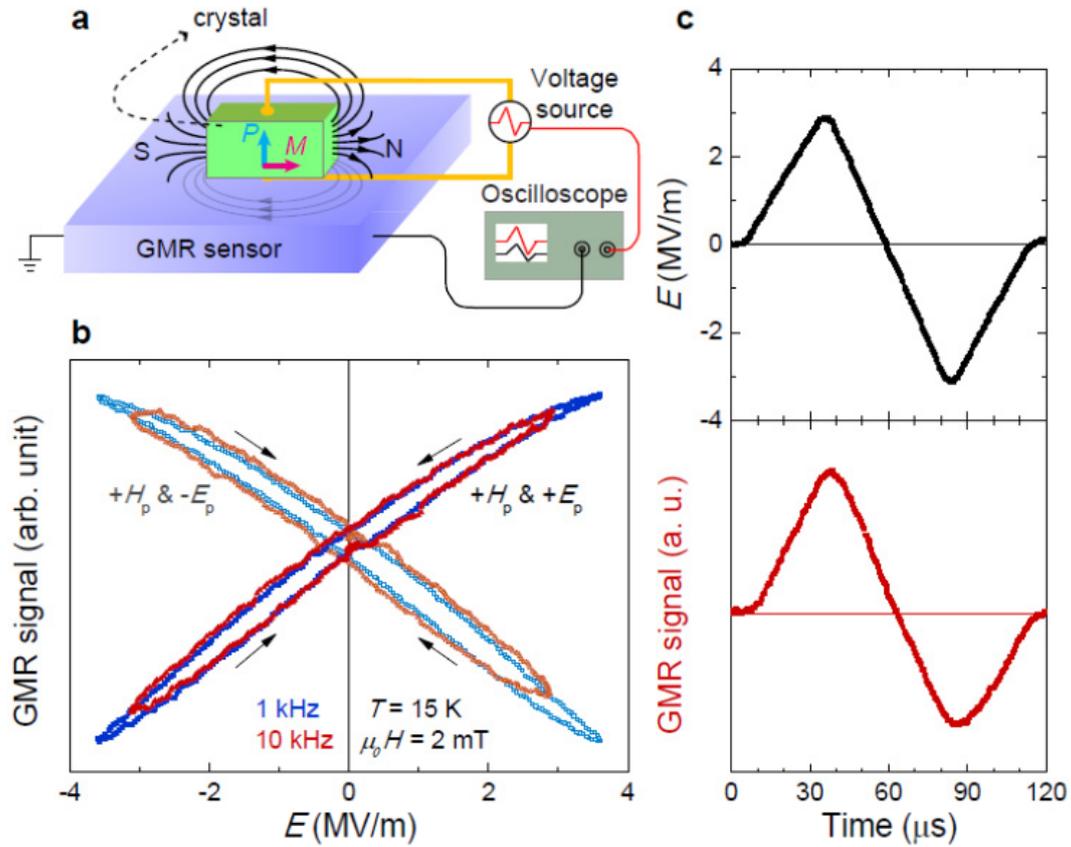

**Figure 5**. **Measurement of fast magnetization change using giant magnetoresistance (GMR) sensor.** (**a**), Schematic drawing of the single crystal bonded on top of a GMR sensor. Electric pulses were applied sequentially while an oscilloscope monitors the time dependence of the electric pulse and the voltage change from the GMR sensor. (**b**), $E$-dependence of the GMR signals at 15 K. The maximum modulation of the GMR signals under the electric field sweep corresponds to the equivalent magnetic field changes of about 2.5 mT. Note that the hysteresis in the GMR signal vs. electric field graph mainly comes from the changes in the remanent $M$. (**c**), The full increase or decrease of the GMR signals occurs within ~30 μs, limited by the rise or fall time of the electric pulses, indicating that the magnetization jumps can occur at least in much less than 30 μs.

**Supplementary Information**

**Determination of the ferroelectric phase diagrams of the optimally O$_2$ atmosphere annealed Ba$_{0.5}$Sr$_{1.5}$Zn$_2$(Fe$_{0.92}$Al$_{0.08}$)$_{12}$O$_{22}$ crystals:** For Y-type multiferroic hexaferrites, the oxygen annealing can change a bit the magnetic and ferroelectric phase boundaries[27]. Therefore, we determined the ferroelectric phase diagram of the studied crystal of which magnetization vs. temperature indicated the long range antiferromagnetic ordering at $T_N$ = ~260 K. It is known that one can determine the ferroelectric phase diagram by measuring the dielectric anomaly (peak) as a function of magnetic field in the isothermal condition[30]. By mapping out the peak field positions in dielectric constant vs. magnetic field plots at selected temperatures, we plotted the ferroelectric phase diagram of the annealed crystals, as summarized in Fig. S1. From the study, we found that the dielectric constant peaks can coin to the switching from the paraelectric to ferroelectric phase boundary but can also appear at zero magnetic field upon having field-induced ferroelectric domain switching. Inside the transverse conical state stabilized at zero $H$ bias after the ME after processing, we found a characteristic peak at zero magnetic field due to the ferroelectric domain switching up to around 150 K, in addition to the high field peaks located above 1 tesla for crossing the paraelectric phase boundary.

At around 150 K, the low field peak has increased a bit in magnetic field toward 10–20 mT scale and then at higher temperature the peak structure became broadened, indicating that the transverse conical state is only stable below 150 K at zero field. From the neutron diffraction study[15], we found above 150 K that alternating longitudinal conical state is stable up to around ~230 K at zero magnetic field. Between $T_N$ and 230 K, there seems to be a mixed phase between a modified helix and alternating longitudinal conical states. Therefore, above ~150 K, there seems no ferroelectric phase at zero magnetic field and only a field induced crossover from paraelectric (longitudinal conical or modified helix phase) to ferroelectric (transverse conical) phase seems to be gradually appearing.

**Weak magnetic in-plane anisotropy in BSZFA:** The transverse conical spin configuration in the BSZFA crystal has magnetic anisotropy of easy plane type. This feature can be proved by handling the cone axis with small rotating magnetic field of several tens of mT[31]. We demonstrated the angle-dependence of $P$ under the rotating $H$ of 20 mT, 0.2 T, and 1 T, respectively, applied in the *ab* plane normal to [001] (horizontal rotation) at 15 K, as shown in Fig. S2. All the *P* values follow nearly sinusoidal curves, strongly supporting that the cone axis should follow the direction of *H* while keeping the spin helicity intact. The highly tunable *P* under *H* as small as 20 mT highlights the weak in-plane anisotropy inside each domain of the transverse conical states with minimal anisotropy along the hexagonal axes within the plane.

**Supplementary References**

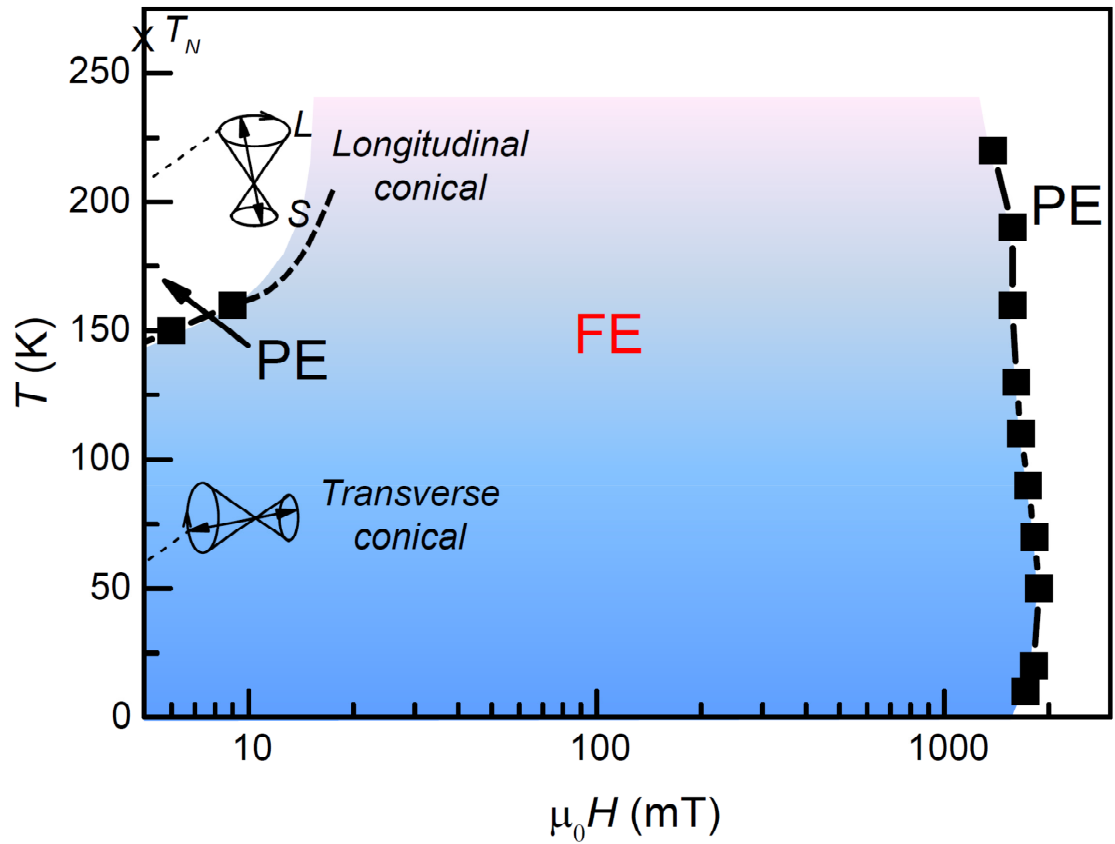

**Figure S1. The ferroelectric phase diagrams of the optimally O$_2$ atmosphere annealed Ba$_{0.5}$Sr$_{1.5}$Zn$_2$(Fe$_{0.92}$Al$_{0.08}$)$_{12}$O$_{22}$ crystals.** The phase boundaries were derived from the peak positions of $H$-dependent dielectric constant in the $H$-decreasing sweep. The transverse conical state at zero magnetic field seems to be stabilized below ~150 K.

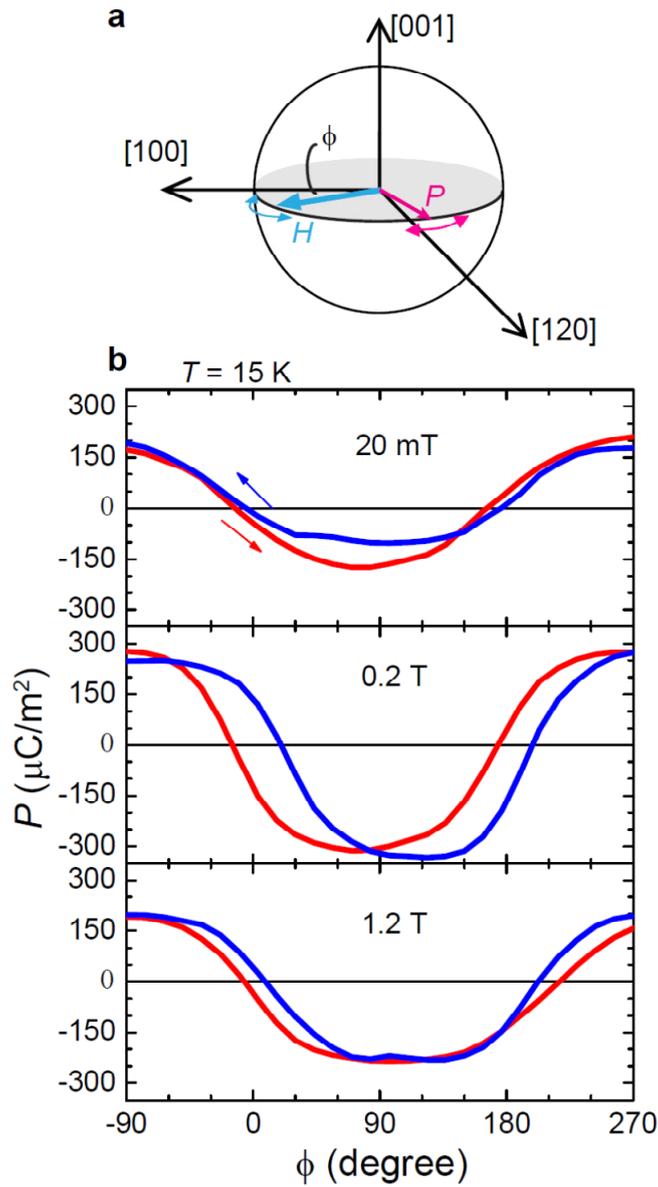

**Figure S2**. **Angle dependence of *P* under the horizontally rotation of *H* at 15 K**. (**a**), Schematic configurations for the measurements of electric polarization under the horizontally rotating *H*. ϕ is defined as the relative angles between *H* and [100] direction. (**b**), ϕ dependence of the polarization under the rotating *H* of 20 mT, 0.2 T, and 1 T. These results support that the FC single ferrimagnetic/ferroelectric domains (at a small bias field of 20 mT) holds the orthogonal relationship with each other, and rotates smoothly with rotating in-plane magnetic fields inside each domain with minimal anisotropy along the hexagonal axes within the plane.